\documentclass[showpacs,aps,prc,preprint,nofootinbib,superscriptaddress]{revtex4}

\usepackage[utf8]{inputenc}
\usepackage{graphicx,subfigure}
\usepackage{float}
\usepackage{amsmath}
\usepackage{amssymb}
\usepackage{color}
\usepackage{slashed}

\newcommand{\bea}{\begin{eqnarray}}
\newcommand{\eea}{\end{eqnarray}}
\newcommand{\be}{\begin{eqnarray}}
\newcommand{\ee}{\end{eqnarray}}

\begin{document}


\title{Solution of n-$^4$He elastic scattering problem using Faddeev-Yakubovsky equations}


\author{Rimantas Lazauskas}
\email[]{rimantas.lazauskas@iphc.cnrs.fr} \affiliation{
Universit{\'e} de Strasbourg, CNRS, IPHC UMR 7178, F-67000
Strasbourg, France}




\date{\today}

\begin{abstract}
The first ever numerical solution of  five-body Faddeev-Yakubovsky equations
 is presented in this work.
Modern realistic Nucleon-Nucleon Hamiltonians have been tested
when describing low energy elastic neutron scattering on $^4$He
nucleus. Obtained results have been  compared with those available
in the literature and  based on solution of the Schr\"{o}dinger
equation.

\end{abstract}


\maketitle


 Solution of nuclear bound
state problem by ab-initio methods have reached new heights during
the last decade~\cite{Pieper2002,Hagen:2010prc,Barrett:2013ppnp,IMSRG,Gorkov_barb}.
 Accurate description of the nuclei composed of several
nucleons has been made possible. However bound-state properties,
such as binding energies, nuclear densities and radii, provide
only a rather restricted set of data with which to test our
understanding  of the nuclear force.  It is the nuclear scattering  
experiment, where cross sections can be measured  as a function of
energy, reaction channel, angular distributions, and polarization
phenomena, provides the richest set of data on nuclear interaction
and dynamics.

However description of a few-nucleon scattering problem in its
full complexity turns to be quite problematic. The main difficulty
is related with the fact that unlike bound-state wave function,
which asymptotically approaches zero for large values of any
two-particle separation, the scattering wave function is
not-compact. When solving a scattering problem in configuration
space proper treatment of boundary conditions is required, which
in a Lippmann-Schwinger equation formulation of the scattering
problem are ill defined. In early 60's Faddeev formulated the
t-matrix approach to the three-body problem~\cite{Faddeev:1960su},
providing a proper way to formulate boundary conditions for
continuum problems dominated by the short-ranged interactions.
Just a few years later Faddeev's revolutionary work has been
generalized to any number of particles by
Yakubovsky~\cite{yakubovsky:67en}. Regardless these revolutionary
mathematical developments progress in solution of
Faddeev-Yakubovsky equations (FYe) is slow and for long years was
limited to A=3 and A=4
cases~\cite{Glockle_bible,PhysRevC.95.034003}. The main difficulty
is related to the complexity of these equations. Indeed in
Faddeev-Yakubovsky (FY) approach a few-particle Schr\"{o}dinger
equation is transformed into a set of differential equations for
so called FY components, which are introduced with a purpose to
uncouple asymptotes of the binary  scattering channels. The number
of these components (channels) increases like a factorial of a
particle number, resulting into very poor scaling of FY formalism
with a particle number.

One should mention that FY equations is not an unique way to solve
scattering problem in configuration space. Diverse scattering
problems may be solved accurately also based on the
Schr\"{o}dinger equation, if the Faddeev decomposition (or its
equivalent) is used in order to enforce the proper boundary
conditions~\cite{Viviani_4B}. Furthermore, due to poor scaling of
FYe's with a particle number, approaches based on Schr\"{o}dinger
equation, like~\cite{Navratil_prc82:2010}, have much brighter
prospects than FY approach in describing systems containing more
than five particles. However once addressing scattering problem
with a Schr\"{o}dinger equation one should be cautious about the
possibility to end up with the spurious solutions. Therefore if
computationally accessible, due to their mathematically rigorous
nature FYe formalism remains a reference in solving few-particle
scattering problem.

In this study the first solution of FYe in configuration
space is presented for a five-body system. Modern realistic
Nucleon-Nucleon interactions will be employed to describe neutron
elastic scattering on $^4$He nucleus. Results will be compared
with those available in the literature and obtained using methods
based on solving Schr\"{o}dinger equation.

Calculations  have been performed for three significantly
different realistic nucleon-nucleon interaction models. The
considered potentials describe very accurately NN scattering data
and include the tail parts determined by the pion-exchange between
the nucleons. Nevertheless these models differ significantly in
the procedure adapted to parameterize their short range parts. The
AV18 model is a local NN potential~\cite{AV18}; INOY04 model
contains strongly non-local core within R=2~fm range for S and P
waves~\cite{Dole_PRC_04}; whereas I-N3LO potential~\cite{IN3LO_03}
is non-local in momentum space and is based on the $\chi$EFT
approach, being derived
 up to next-to-next-to-next-to-leading
order in chiral perturbation theory.
 All the results presented in what follows have been
obtained considering equal masses for neutrons and protons ($m_n=m_p=m$)
with ${\frac{\hbar^2}{m}}=41.471$ MeV fm$^2$.

\section{Formalism}
\section{5-body FY equations}

\begin{figure}[h!]
\begin{center}
\includegraphics[width=0.75\textwidth]{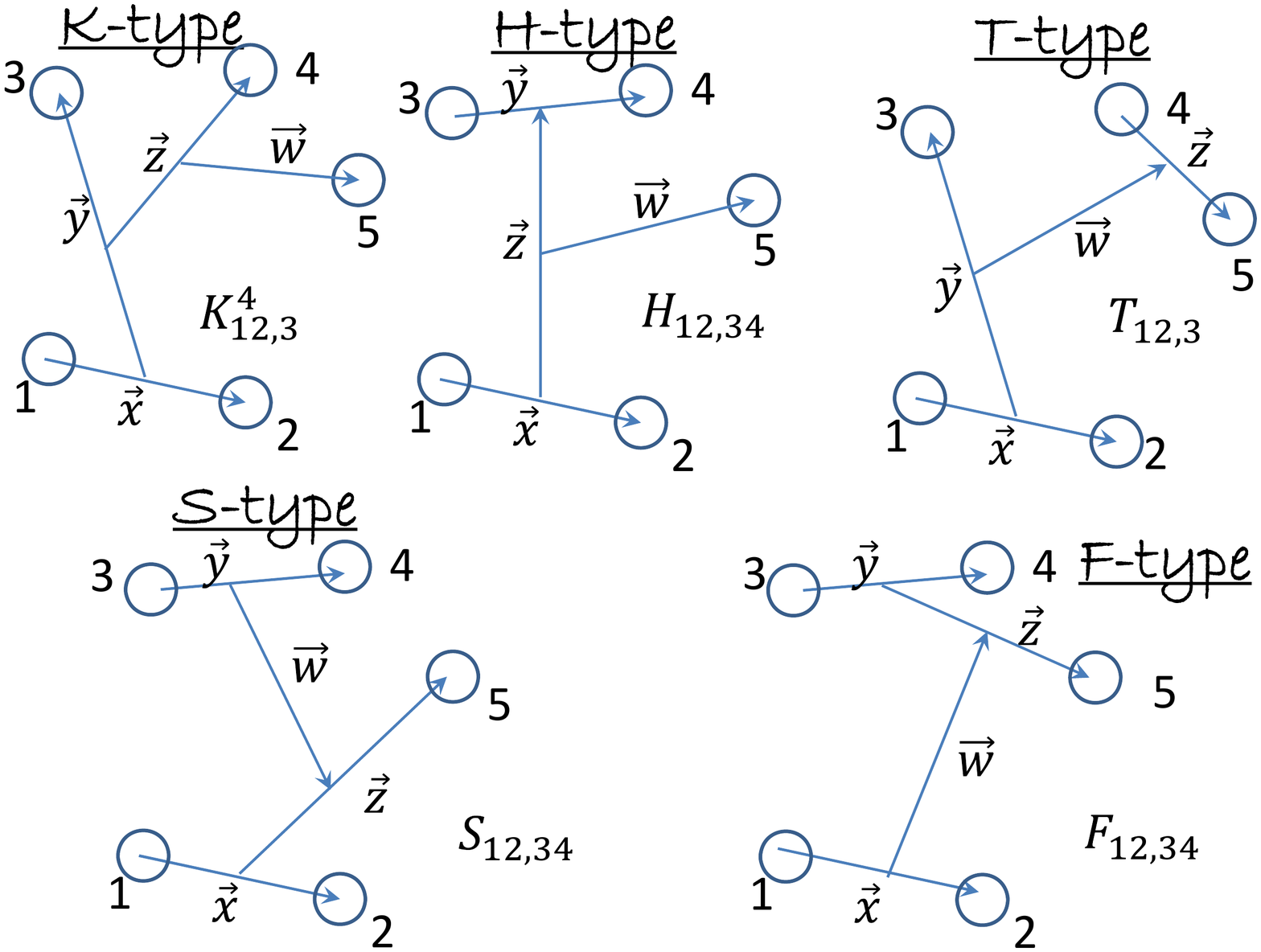}
\end{center}
\caption{ (Color online) 5-particle Jacobi coordinate sets used to
describe FY components, denoted in this work as $K,T,H,S,F$.}
\label{Fig_5b_config}
\end{figure}

In the late sixties Yakubovsky demonstrated a scheme how to
generalize 3-body Faddeev equations to N-body system of particles
governed by short-ranged interactions~\cite{yakubovsky:67en}. A detailed derivation of
5-body FY equations has been performed in ref.~\cite{Sasakawa_5b}. Derivations of
N-body FY\ equations starts by decomposing systems total wave
function into binary partitions, similar to 3-body
Faddeev components:%
\begin{eqnarray}
\phi _{ij} =G_{0}V_{ij}\Psi.
\end{eqnarray}
For a five body system one may construct 10\ different binary
components  by permuting particle indexes $(ij)$. In that follows
I will denote by letters $(ijklm)$ combination of particle indexes
$(12345)$. It is easy to verify that the total systems wave
function is recovered by simply adding binary components:
\begin{eqnarray}
\Psi (x,y,z,w) =\sum\limits_{i<j}^{5}\phi _{ij}(x,y,z,w).
\end{eqnarray}%
The binary components $\phi _{ij}(x,y,z,w)$ are further split into
four-body type components by following a pattern of breaking
five body $(ijklm)$ partition into clusters and their subclusters.
One has two types of four-body
components, which are similar to ones appearing in
four-body FY equations:
\begin{equation}
\left\{
\begin{array}{c}
\psi _{ij}^{ijk}=G_{ij}V_{ij}(\phi _{jk}+\phi _{ki}),\qquad  \\
\psi _{ij}^{ij,kl}=G_{ij}V_{ij}\phi _{kl}.%
\end{array}%
\right.
\end{equation}%
Here 5-body Green's $G_{ij}$ includes single interaction term
$V_{ij}$, i.e. $G_{ij}=(E-H_0-V_{ij})^{-1}$. For a 5-body system
there exist 30 different 4-body components of type $\psi
_{ij}^{ijk}$ as well as 30 components of type $\psi
_{ij}^{ij,kl}$. Using Yakubovsky's scheme one may easily decompose
the binary components into the 4-body ones:
\begin{equation}
\phi _{ij}=\psi _{ij}^{ijk}+\psi _{ij}^{ijl}+\psi _{ij}^{ijm}+\psi
_{ij}^{ij,kl}+\psi _{ij}^{ij,km}+\psi _{ij}^{ij,lm}.
\end{equation}%
Finally, the four-body components are decomposed into a sum of
5-body FY components:
\begin{equation}
\left\{
\begin{array}{cc}
\psi _{ij}^{ijk}=K_{ij,k}^{l}+K_{ij,k}^{m}+\mathcal{T}_{ij,k}\qquad  & \text{%
(30 amplitudes)}, \\
\psi _{ij}^{ij,kl}=H_{ij,kl}+\mathcal{S}_{ij,kl}+\mathcal{F}_{ij,kl} & \text{%
(30 amplitudes)}.%
\end{array}%
\right.
\end{equation}%
Faddeev-Yakubovsky equations involve five different types of
5-body FY components (see Fig.\ref{Fig_5b_config}), denoted in
this work by $K_{ij,k}^{l},T_{ij,k,}H_{ij,kl},S_{ij,kl}$ and
$\mathcal{F}_{ij,kl}$. In total, there exist 60 components of
type-K and 30 components for each of the types $H,S,T$ and
$\mathcal{F}$. 5-body FY equations constitutes a set of 180
coupled equations, each of which might be associated with a
particular FY component. One has 5 non-trivial equations, each
highlighting one particular component of different type. By
separating terms associated with a highlighted component in the
right hand side of the relations -- these equations can be
summarized as follows:
\begin{eqnarray}
(E-\widehat{H}_{0}-V_{12})K_{12,3}^{4}
&=&V_{12}(K_{13,2}^{4}+K_{23,1}^{4}+K_{13,4}^{5}+K_{23,4}^{5}+K_{13,4}^{2}+K_{23,4}^{1}
\notag\\
&&+\mathcal{T}_{13,4}+\mathcal{T}_{23,4} \notag\\
&&+H_{13,24}^{{}}+H_{23,14}^{{}}+\mathcal{S}_{13,24}^{{}}+\mathcal{S}%
_{23,14}^{{}}+\mathcal{F}_{13,24}^{{}}+\mathcal{F}_{23,14}^{{}}),\notag \\
(E-\widehat{H}_{0}-V_{12})H_{12,34} &=&V_{12}(H_{34,12}\notag \\
&&+K_{34,1}^{2}+K_{34,2}^{1}+K_{34,1}^{5}+K_{34,2}^{5} \notag\\
&&+\mathcal{T}_{34,1}+\mathcal{T}_{34,2}), \notag\\
(E-\widehat{H}_{0}-V_{12})\mathcal{T}_{12,3} &=&V_{12}(\mathcal{T}_{13,2}+%
\mathcal{T}_{23,1}  \\
&&+H_{13,45}+H_{23,45}\notag \\
&&+\mathcal{S}_{13,45}+\mathcal{S}_{23,45}\notag \\
&&+\mathcal{F}_{13,45}+\mathcal{F}_{23,45}), \notag\\
(E-\widehat{H}_{0}-V_{12})\mathcal{S}_{12,34} &=&V_{12}(\mathcal{F}_{34,12}+%
\mathcal{S}_{34,15}+\mathcal{S}_{34,25}\notag \\
&&+\mathcal{F}_{34,15}+\mathcal{F}_{34,25} \notag\\
&&+H_{34,15}+H_{34,25}), \notag\\
(E-\widehat{H}_{0}-V_{12})\mathcal{F}_{12,34}
&=&V_{12}(\mathcal{S}_{34,12} \notag
\\
&&+K_{34,5}^{1}+K_{34,5}^{2} \notag\\
&&+\mathcal{T}_{34,5}). \notag \label{eq_5b_fy}
\end{eqnarray}%
Other equations follow from this set by simply permuting particle
indexes in the ordered way. For a system of five identical
particles one can reduce the problem to solving only one set of
the five equations, since there remains only five
independent FY components ($K_{ij,k}^{l},T_{ij,k,}H_{ij,kl},S_{ij,kl}$ and $%
\mathcal{F}_{ij,kl}$). Other components may be obtained from a
selected set of components ($K,H,S,T,\mathcal{F}$) by using
particle-permutation symmetry relations.

\section{Coordinates}

Each FY component $F=(K,T,H,S,\mathcal{F})$ is a function in
twelve-dimensional configuration space, determined by the four 3D
vectors ($\vec{x},\vec{y},\vec{z},\vec{w}$). It is convenient to
express the FYCs in their proper set of Jacobi coordinates, see
Fig.~\ref{Fig_5b_config}. Jacobi coordinate connecting two
clusters $\left( s\right) $ and $\left( t\right) $ are expressed
using a general formulae:
\begin{equation}
(\overrightarrow{x},\overrightarrow{y},\overrightarrow{z},\overrightarrow{w}%
)=\sqrt{\frac{2m_{s}m_{t}}{m(m_{s}+m_{t})}}\left( \overrightarrow{r}_{s}-%
\overrightarrow{r}_{t}\right) ,
\end{equation}%
where $m_{s}$ and $m_{t}$ are the masses of the clusters, while $%
\overrightarrow{r}_{s}$ and $\overrightarrow{r}_{t}$ are
respective positions of their center-of-masses. A mass factor $m$
of free choice is introduced into the former expression in order
to retain the proper units of the distances. When studying systems
of identical particles it is convenient to identify this mass with
the mass of a single particle (in this study mass of a nucleon).
In terms of Jacobi coordinates center-of-mass free Hamiltonian is
expressed:
\begin{equation}
H_{0}=-\frac{\hbar ^{2}}{m}\left( \Delta _{x}+\Delta _{y}+\Delta
_{z}+\Delta _{w}\right).
\end{equation}
When studying low energy processes partial wave expansion turns to be a
very
efficient tool to express angular dependence of the systems wave
function. Without exception in this work partial wave expansion is
used to describe spatial dependence of FY components as well as
their dependence on spin and isospin quantum numbers:
\begin{equation}
F^{JM}(\overrightarrow{x},\overrightarrow{y},\overrightarrow{z},\overrightarrow{w}%
)=\sum \frac{f_{\alpha }(x,y,z,w)}{xyzw}\left| \left\{ \left\{
l_{x}l_{y}\right\} _{l_{xy}}\left\{ l_{z}l_{w}\right\}
_{l_{zw}}\right\} _{L}\left\{ S\right\} \right\rangle _{JM}\left\{
T\right\} _{TT_{z}},
\label{eq_pw_exp}
\end{equation}%
here $\alpha \equiv
(l_{x},l_{y},l_{z},l_{w},l_{xy},l_{zw},L,\left\{ S\right\}
,\left\{ T\right\} )$ is an index representing set of intermediate
quantum numbers, coupled to total angular momentum $J$ and total
isospin $T$ with its projection $T_{z}$ (for a n-$^4$He scattering
considered in this work, total isospin and its projection are
fixed to T=1/2 and T$_z$=1/2$^{-}$. In the last expression $\left\{
S\right\} $ and $\left\{ T\right\} $ represent respectively
partial-wave basis\ dependence on spin and isospin, which is provided
by:
\begin{equation}
\left\{ S\right\} =\left| \left\{ \left\{ s_{1}s_{2}\right\}
_{s_{x}}\left\{ s_{3}s_{4}\right\} _{s_{y}}\right\}
_{s_{xy}}s_{5}\right\rangle _{SS_{Z}},
\label{eq_spin_bs}
\end{equation}%
here $s_{1}$ to $s_{5}$ are spins of individual nucleons, whereas $%
s_{x},s_{y},s_{xy},S$ represent quantum numbers of intermediate
couplings. An equivalent expression is used to develop isospin dependence
$\left\{ T\right\} $ of FY components. The reduced components $f_{\alpha}(x,y,z,w)$ represent
dependence on radial parts of the coordinates. This dependence is
expressed using Lagrange-Laguerre basis functions.

The last set of equations~(\ref{eq_pw_exp}-\ref{eq_spin_bs})
define the principal partial-wave basis set employed in this work.
However, in parallel, two additional equivalent partial wave
coupling schemes have been used. One exposing coupling of angular
momenta $\left\{ l_{y}l_{z}\right\} _{l_{yz}}$, required in order
to perform some permutation operations in yz-space, another
explicitly exposing two-particle angular momentum
$\left\{l_x(s_1,s_2)_{s_x}\right\}_{j_x}$ used to evaluate matrix
elements of NN-interaction between particles 1 and 2.

\section{Operators}

In order to solve FY equations it is useful to define a set of
operators, which allow to couple different FY components. First, I
introduce a group of operators which couple FY components of
different type, but which share the same particle ordering:
\begin{eqnarray}
K_{12,3}^{4} &=&\left( P^{KH}\right) _{yz}^{1}H_{12,34};\quad
H_{12,34}=\left( P^{HK}\right) _{yz}^{1}K_{12,3}^{4}, \notag \\
K_{12,3}^{4} &=&\left( P^{KT}\right)
_{zw}^{1}\mathcal{T}_{12,3};\quad
\mathcal{T}_{12,3}=\left( P^{TK}\right) _{zw}^{1}K_{12,3}^{4},\quad \notag \\
H_{12,34} &=&\left( P^{HS}\right)
_{zw}^{1}\mathcal{S}_{12,34};\quad
\mathcal{S}_{12,34}=\left( P^{SH}\right) _{zw}^{1}H_{12,34}, \notag \\
H_{12,34} &=&\left( P^{HF}\right)
_{zw}^{1}\mathcal{F}_{12,34};\quad
\mathcal{F}_{12,34}=\left( P^{FH}\right) _{zw}^{1}H_{12,34},  \\
\mathcal{S}_{12,34} &=&\left( P^{SF}\right) _{zw}^{1}\mathcal{F}%
_{12,34};\quad \mathcal{F}_{12,34}=\left( P^{FS}\right) _{zw}^{1}\mathcal{S}%
_{12,34}, \notag \\
\mathcal{T}_{12,3} &=&\left( \underline{P}^{TS}\right) _{yz}^{0}\mathcal{S}%
_{12,45};\quad \mathcal{S}_{12,34}=\left( \underline{P}^{ST}\right) _{yz}^{0}%
\mathcal{T}_{12,5}, \notag \\
\mathcal{S}_{34,12} &=&\left( \underline{P}^{SF}\right) _{xy}^{0}\mathcal{F}%
_{12,34};\quad \mathcal{F}_{34,12}=\left( \underline{P}^{FS}\right) _{xy}^{0}%
\mathcal{S}_{12,34}. \notag
\end{eqnarray}
Operators presented on each line represent inverse to each other,
i.e. as example $\left(\left( P^{HK}\right)
_{yz}^{1}\right)^{-1}=\left( P^{HK}\right)_{yz}^{1}$. Expressions
of these operators splits in to  tensor product of operators
acting in coordinate, spin and isospin spaces. When matrix
elements of these operators are properly ordered inverse operator
is directly obtained from the original operator by simply
permuting its matrix elements and thus does not require separate
evaluation or storage.

Second group of operators is used to change the particle ordering:
\begin{eqnarray}
K_{12,3}^{4} &=&\left( P^{+}\right) _{xy}^{1}K_{23,1}^{4};\quad
K_{12,3}^{4}=\left( P^{-}\right) _{xy}^{1}K_{31,2}^{4},  \notag \\
\mathcal{T}_{12,3} &=&\left( P^{+}\right)
_{xy}^{1}\mathcal{T}_{23,1};\quad
\mathcal{T}_{12,3}=\left( P^{-}\right) _{xy}^{1}\mathcal{T}_{31,2}, \\
K_{12,3}^{4} &=&\left( \varepsilon P^{34}\right)
_{yz}^{1}K_{12,4}^{3},
\notag \\
K_{12,3}^{5} &=&\left( \varepsilon P^{45}\right)
_{zw}^{1}K_{12,3}^{5},
\notag \\
H_{12,34}^{{}} &=&\left( P^{H}\right) _{xy}^{0}H_{34,12}^{{}} , \notag \\
\left( \widetilde{P}^{3}\right) _{xy}^{1} &=&\left( P^{+}\right)
_{xy}^{1}+\left( P^{-}\right) _{xy}^{1},  \notag \\
\left( \widetilde{P}^{4}\right) _{yz}^{1} &=&\left( \varepsilon
P^{34}\right) _{yz}^{1};\ \quad \left( \widetilde{P}^{5}\right)
_{zw}^{1}=\left( \varepsilon P^{45}\right) _{zw}^{1}. \notag
\end{eqnarray}%
In these expressions operators have been denoted using the general
notation $\left( P^{A}\right)_{xy}^{n}$, where integer $n$
indicates number of angular integrations involved in coupling
partial amplitudes; $xy$- denotes that this operator transforms
radial dependencies of the amplitude in coordinates $x$ and $y$.
Expressions for these operators are quite trivial, equivalent to
ones used in solving 3-body or 4-body FY equations. Nevertheless
their expressions become quite voluminous and will be published
elsewhere. When applied successively, this set of operators is
sufficient to couple any two FY components and thus solve 5-body
FY equations as formulated in eq.(\ref{eq_5b_fy}). Using these
definitions 5-body FY equations read:
\begin{eqnarray}
K_{12,3}^{4} &=&G_{12}V_{12}\left( \widetilde{P}^{3}\right)
_{xy}^{1}\left( K_{12,3}^{4}+\left( P^{KH}\right) _{yz}^{1}\left[
H_{12,34}^{{}}+\left( P^{HS}\right)
_{zw}^{1}\mathcal{S}_{12,34}^{{}}+\left( P^{HF}\right)
_{zw}^{1}\mathcal{F}_{12,34}^{{}}\right] \right. ,   \notag \\
&&\left. +\left( \widetilde{P}^{4}\right) _{yz}^{1}\left[ K_{12,3}^{4}+%
\left( \widetilde{P}^{5}\right) _{zw}^{1}K_{12,3}^{4}+\left(
P^{KT}\right)
_{zw}^{1}\mathcal{T}_{12,3}\right] \right) , \\
H_{12,34} &=&G_{12}V_{12}\left( \underline{P}^{H}\right)
_{xy}^{0}\left( H_{12,34}+2\left( P^{HK}\right) _{yz}^{1}\left[
K_{12,3}^{4}+\left(
\widetilde{P}^{5}\right) _{zw}^{1}K_{12,3}^{4}+\left( P^{KT}\right) _{zw}^{1}%
\mathcal{T}_{12,3}\right] \right),   \notag \\
\mathcal{T}_{12,3} &=&G_{12}V_{12}\left( \widetilde{P}^{3}\right)
_{xy}^{1}\left( \mathcal{T}_{12,3}+\left( \underline{P}^{TS}\right) _{yz}^{0}%
\left[ \mathcal{S}_{12,34}+\left( P^{SF}\right) _{zw}^{1}\mathcal{F}%
_{12,34}+\left( P^{SH}\right) _{zw}^{1}H_{12,34}\right] \right),   \notag \\
\mathcal{S}_{12,34} &=&G_{12}V_{12}\left( \left(
\underline{P}^{ST}\right)
_{yz}^{0}\left( \widetilde{P}^{3}\right) _{xy}^{1}\left( \underline{P}%
^{TS}\right) _{yz}^{0}\left( \underline{P}^{SF}\right)
_{xy}^{0}\left[ \mathcal{F}_{12,34}+\left( P^{FH}\right)
_{zw}^{1}H_{12,34}+\left(
P^{FS}\right) _{zw}^{1}\mathcal{S}_{12,34}\right] \right. ,  \notag \\
&&+\left. \left( \underline{P}^{SF}\right) _{xy}^{0}\mathcal{F}%
_{12,34}\right),   \notag \\
\mathcal{F}_{12,34} &=&G_{12}V_{12}\left( \underline{P}^{FS}\right) _{xy}^{0}%
\left[ \mathcal{F}_{12,34}+\left( \underline{P}^{ST}\right)
_{yz}^{0}\left( \mathcal{T}_{12,3}+2\left( P^{TK}\right)
_{zw}^{1}K_{12,3}^{4}\right) \right]. \notag
\label{eq_5b_op_form}
\end{eqnarray}
Since in this work a system of five formally identical particles is
considered, the last set of equations is written for the
components, where particles are ordered in a natural succession
(12345).

The last set of equations is sufficient to solve 5-body problem
and obtain to this problem related physical observables -- binding
energies or phaseshifts. However in order to estimate expectations
values of the physical operators one may require to reproduce total
systems wave function, which may be expressed in terms of FY
components:
\begin{eqnarray}
\Psi (x,y,z,w) &=&\sum\limits_{i<j}^{5}\phi _{ij}(x,y,z,w) \notag \\
\phi _{12} &=&\psi _{12}^{123}+\psi _{12}^{124}+\psi
_{12}^{125}+\psi
_{12}^{12,34}+\psi _{12}^{12,35}+\psi _{12}^{12,45} \\
&=&\left[ 1+P^{34}+P^{45}P^{34}\right] \left( \psi
_{12}^{123}+\psi _{12}^{12,34}\right), \notag
\end{eqnarray}
now we denote:
\begin{eqnarray}
\widetilde{X} &=&X+R_{X},\notag \\
X &\equiv &(K,H,\mathcal{T},\mathcal{S},\mathcal{F}\mathbb{)},
\end{eqnarray}
and where the term $R_{X}$ represents a sum of components
appearing on the right-hand side of the FY equation
(\ref{eq_5b_fy}) relevant to component X. For example:
\begin{eqnarray}
X &=&K_{12,3}^{4}, \notag \\
R_{x}
&=&K_{13,2}^{4}+K_{23,1}^{4}+K_{13,4}^{5}+K_{23,4}^{5}+K_{13,4}^{2}+K_{23,4}^{1}+
\notag \\
&&+\mathcal{T}_{13,4}+\mathcal{T}_{23,4} \\
&&+H_{13,24}^{{}}+H_{23,14}^{{}}+\mathcal{S}_{13,24}^{{}}+\mathcal{S}%
_{23,14}^{{}}+\mathcal{F}_{13,24}^{{}}+\mathcal{F}_{23,14}^{{}},
\end{eqnarray}
then
\begin{equation}
\begin{array}{c}
\widetilde{\psi }_{12}^{123}=\widetilde{K}_{12,3}^{4}+\widetilde{K}%
_{12,3}^{5}+\widetilde{\mathcal{T}}_{12,3}=\left(
1+P^{45}\right) \widetilde{K}_{12,3}^{5}+\widetilde{\mathcal{T}}%
_{12,3},\qquad \\
\widetilde{\psi }_{12}^{12,34}=\widetilde{H}_{12,34}^{{}}+\widetilde{%
\mathcal{S}}_{12,34}^{{}}+\widetilde{\mathcal{F}}_{12,34}^{{}}.%
\end{array}%
\end{equation}
Finally
\begin{equation}
\Psi =\left[ 1+\left( 1+P^{45}\right) P^{34}\right] \left( \widetilde{\psi }%
_{12}^{123}+\widetilde{\psi }_{12}^{12,34}\right).
\end{equation}
\subsection{Boundary conditions}

Solution of the differential equations is not complete, unless
proper boundary conditions are formulated and imposed. The reduced
components are regular functions both when related to the solution
of bound state or scattering problems,
\begin{equation}
f_{\alpha}(0,y,z,w)=f_{\alpha}(x,0,z,w)=f_{\alpha}(x,y,0,w)=f_{\alpha}(x,y,z,0).
\end{equation}
It is the boundary condition for the asymptotic region (at large
radial distances) turns to be more complicated when a scattering
problem is considered. For a bound state problem FY components are
compact and thus square-integrable basis functions might be
readily used to describe behavior of the reduced components. For
the scattering problems, which does not involve systems
decomposition into more than two clusters (a case considered in
this work), reduced components still remain compact in $x,y,z$
directions. On the other hand asymptotic parts of elastic incoming
(outgoing) wave of the scattered clusters are expressed in
$w$-radial dependence of  the reduced FY components. In order to
fulfill this feat but at the same time to be able to use
square-integrable basis functions in solving scattering problem
the reduced components are split in two terms
\begin{equation}
f_{\alpha,a}(x,y,z,w)=\widetilde{f}^{sh}_{\alpha,a}(x,y,z,w)+\widetilde{f}^{ass}_{\alpha,a}(x,y,z,w).
\end{equation}
In the last expression index $a$ indicates an incoming channel
number, for which solution is searched. The term
$\widetilde{f}^{sh}_{\alpha,a}(x,y,z,w)$  is intended to describe
only interior part of the component $f_{\alpha,a}(x,y,z,w)$  based
on expansion employing compact basis functions. The  term
$\widetilde{f}^{ass}_{\alpha,a}(x,y,z,w)$ complements the
expression in order to describe properly asymptotic part of the
reduced FY components. This term takes a form
\begin{equation}
\widetilde{f}^{ass}_{\alpha,a}(x,y,z,w)=\sum_{b}\sum_{\beta\subset
b}
\delta_{\beta,\alpha}\widetilde{\phi}_{\beta}(x,y,z)\left(\delta_{a,b}\hat{j}_{l^\alpha_w}(q_{b}w)+\sqrt{\frac{q_a}{q_b}}K_{b,a}\hat{n}_{l^\alpha_w}(q_{b}w)
\eta_{l^\alpha_w}^{reg}(w)\right).
\label{eq_f_ass}
\end{equation}
In the last expression the first sum runs over all open channels $b$, whereas the
second sum runs over all
the partial-wave amplitudes $\beta\subset b$, contributing in
expanding asymptotes of this channel. The term $K_{b,a}$ represent
the K-matrix elements, describing scattering process, to be
determined. The $\hat{j}_{l^\alpha_w}(q_{b}w)$ and $\hat{n}_{l^\alpha_w}(q_{b}w)$ represent respectively
Riccati-Bessel and Riccati-Neumann functions. Additionally a
function $\eta_l^{reg}(w)$ is introduced in order to regularize diverging behavior
 of Riccati-Neumann function at the origin.  This regularization function
is chosen in a form popularized by the numerical calculations of the
Pisa group~\cite{Viviani_4B,Barletta_PRL103_2009,Kievsky_PRC81_2010}
\begin{equation}
\eta_l^{reg}(w)=\left[1-exp(w/w_0)\right]^{2l+k},
\end{equation}
in this parametrization, the power $k$ parameter  must be chosen
to be $k\geq1$, whereas values $k=1$ and $k=2$ turns to be
optimal. The range parameter $w_0$ draws the matching region
between dominance of $\widetilde{f}^{sh}_{\alpha,a}$ and
$\widetilde{f}^{ass}_{\alpha,a}$ terms and is chosen in the
interval $w_0=(1,2)$ fm. The selected regularization function satisfies natural
conditions
\begin{equation}
\left\{
\begin{array}{c}
\left.\eta_l^{reg}(w)\hat{n}_{l}(q_{b}w)\right|_{w\rightarrow0}=0,
\\
\left.\eta_l^{reg}(w)\hat{n}_{l}(q_{b}w))\right|_{w\rightarrow\infty}=\hat{n}_{l}(q_{b}w).
\end{array}
\right.
\end{equation}
Calculated K-matrix elements to high order turn to be independent
of the two parameters encoded in $\eta_l^{reg}(w)$. This feat
constitutes one of the tests for the reliability of the calculations.

Finally, functions $\widetilde{\phi}_{\beta}(x,y,z)$ represent
bound state-like solutions of the reduced 5-body problem to 4-body
case. For a case considered in this work it represents solution of
bound state problem for $^4He$ nucleus. These functions are
obtained by reducing 5-body problem to 4-body one, which requires
simply eliminating w-dependence in~eq.(\ref{eq_5b_op_form}) -- it
is by equating Laplacian operator ($\Delta _{w}$) as well as all
the permutation operators containing w-dependence to zero.

\subsection{Lagrange-mesh method}

The functions $f_{\alpha,a}(x,y,z,w)$, representing
radial dependence of the FY components, are expanded using basis
functions defined by Lagrange-Laguerre mesh method~\cite{Baye_bible}:
\begin{equation}
f_{\alpha,a}(x,y,z,w)=\sum_{i_{x}=1}^{N_{x,l_x)}}\sum_{i_{y}=1}^{N_{y,l_y}}\sum_{i_{y}=1}^{N_{z,l_z}}\sum_{i_{y}=1}^{N_{w,l_w}}C_{
i_{x},i_{y},i_{z},i_{w}}^{\alpha,a}\mathcal{F}^{l_x}_{i_{x}}\left( x/h_{x,
l_x }\right) \mathcal{F}^{l_y}_{i_{y}}\left(
y/h_{y,l_y}\right)\mathcal{F}^{l_z}_{i_{z}}\left(
z/h_{z,l_z}\right)\mathcal{F}^{l_w}_{i_{w}}\left(
w/h_{w,l_w}\right), \label{eq_LM_basis}
\end{equation}%
with the $C_{i_{x},i_{y},i_{z},i_{w}}^{\alpha,a}$ representing
expansion coefficients to be determined. For low energy physics
low angular momenta components turn to dominant, moreover their radial
 shapes often have more complicated
structure than their high-momenta counterparts. Therefore in this
work number of basis functions is chosen as a function of the
partial angular momentum they represent. This number is gradually
reduced when increasing partial angular momentum number, in a manner
similar to the cases of Hypherspherical Harmonics or Harmonic
oscillator basis with a fixed grand angular momentum number. The
coefficients $h_{x,l_x}$ are scaling
parameters for the basis functions defined as
\begin{equation}
\mathcal{F}^{l_x}_{i}(x)=(-1)^{i}c_{i,l_x}\sqrt{\frac{x}{x_{i}(l_x)}}\frac{L^{2l_x+1}_{N_x(l_x)}(x)}{x-x_{i}(l_x)}%
e^{-x/2},
\end{equation}%
In this expression $L^\alpha_{N}(x)$ denotes a $N^{th}$ degree
generalized Laguerre polynomial, with $x_{i}(l_x)$ representing
zeroes of this polynomial. The coefficients $c_{i,l_x}$ are fixed
by imposing basis functions to be orthonormal, namely:
\begin{equation}
\int_{0}^{\infty }\mathcal{F}^{l_x}_{i}(x)\mathcal{F}^{l_x}_{i^{\prime
}}(x)dx=\delta _{ii^{\prime }}.
\end{equation}
Set of differential equations~(\ref{eq_5b_op_form}) is transformed
into a linear algebra problem by first projecting their angular
dependence on partial wave basis, defined by
eqs.~(\ref{eq_pw_exp}-\ref{eq_spin_bs}), and then projecting
radial parts   on Lagrange-Laguerre mesh basis, defined in
eq.~(\ref{eq_LM_basis}). In this way set of linear equations is
obtained to determine unknown expansion coefficients
$C_{i_{x},i_{y},i_{z},i_{w}}^{\alpha,a}$. This set of equations
may be summarized as follows:
\begin{equation}
\left(\hat{H}^{FY}-E\right) C_{i_{x},i_{y},i_{z},i_{w}}^{\alpha,a}=b^{(a)}.
\label{lin_eq_lm}
\end{equation}
Here $(H^{FY}-E)$ represents the kernel of FY equations acting on
the part of wave function's component defined by the term
$\widetilde{f}^{sh}_{\alpha,a}(x,y,z,w)$ and represented by a set
of linear coefficients $C_{i_{x},i_{y},i_{z},i_{w}}^{\alpha,a}$.
Inhomogeneous term $b^{(a)}$ is constructed  by acting with the
FYe kernel on the part of wave function's component defined by the
$\widetilde{f}^{ass}_{\alpha,a}(x,y,z,w)$ term.

 One may refer
to~\cite{Baye_bible,These_Rimas_03} for a more detailed
description of the numerical methods used in this work.

\subsection{Kohn variational principal}

Projection of the FYe on Lagrange-mesh functions, given
by~eq.(\ref{lin_eq_lm}), provides only as many linear equations as
there exist unknown coefficients
$C_{i_{x},i_{y},i_{z},i_{w}}^{\alpha,a}$. However there exist
additional unknowns due to  the presence of the K-matrix elements
($K_{a,b}$) encoded in parametrization of asymptotic parts of FY
amplitudes $\widetilde{f}^{ass}_{\alpha,a}(x,y,z,w)$, see
eq.(\ref{eq_f_ass}). In order to balance the linear algebra
problem the recourse to Kohn variational principle is made.

Information on the scattering matrix is encoded in the asymptote
of the systems wave function but at the same time in the separate
FY components. Therefore there are two independent ways how to
apply Kohn variational principle. The first one represents the
conventional form of Kohn variational principle, relying on the
Wronskian relation combining total wave function and incoming
wave:
\begin{equation}
K_{a,b}=\sqrt{\frac{1}{q_aq_b}}\left(<\psi_{in,b}|(\hat{H}_0^\theta-E)|\Psi^a>-<\Psi^a|(\hat{H}_0^\theta-E)|\psi_{in,b}>\right).
\label{eq_Kohn_vr}
\end{equation}
In this expression wave function $\psi_{in,b}$ represents a free wave of the channel $b$, defined
by FY partial amplitudes:
\begin{equation}
f^{in}_{\alpha,a}(x,y,z,w)=\sum_{b}\sum_{\beta\subset
b}
\delta_{\beta,\alpha}\widetilde{\phi}_{\beta}(x,y,z)\delta_{a,b}\hat{j}_{l^\alpha_w}(q_{b}w).
\label{eq_f_in}
\end{equation}
The next approach is to replace the total systems wave function by
the set of the Faddeev-Yakubovsky components containing non-zero
$\widetilde{f}^{ass}_{\alpha,a}(x,y,z,w)$ term and encompassing
required K-matrix element:
\begin{equation}
K_{a,b}=\sqrt{\frac{1}{q_aq_b}}\left(<\psi_{in,b}|(\hat{H}_0-E)|\Phi^a>-<\Phi^a|(\hat{H}_0-E)|\psi_{in,b}>\right).
\label{eq_fy_vr}
\end{equation}
In principle, the first relation is mathematically more accurate,
up to second order terms in wave functions
perturbation~\cite{Barletta_PRL103_2009,Kievsky_PRC81_2010}.
However evaluation of this expression requires one to produce the
total systems wave function, which involves calculation of
supplemental multidimensional integrals. In this work these
integrals are performed based on Lagrange-mesh approximation used
to expand FY components, which involves relatively small number of
quadrature points. This approximation weights heavily on the
accuracy of the final results. In practice the second relation
(requiring much smaller numerical effort to evaluate)  turns to be
of the similar accuracy as the first one. Comparison of the
K-matrix elements extracted using two different methods
constitutes a critical test for the accuracy of the calculation
and will be discussed in the next section.

\section{Results}

Solution of the 5-body FY equations turns to be extraordinary
numerical task, which challenge our technical capacities. A
careful choice of the parameter space should be made in order to
optimize solution. The key input is related with a choice of the
Lagrange-mesh basis. One of the criteria used to judge on the
proper basis choice is ability to reproduce ground state binding
energies of $^4$He and $^3$H nuclei, employing the same set of
meshes as will be used in n-$^4$He scattering calculations.
Table~\ref{tab:he4_be} resumes the binding energies of $^4$He,
obtained for the parameter space to be employed in n-$^4$He
scattering calculations. The partial wave expansion was
constructed by limiting partial angular momenta to those
satisfying $max(l_x,l_y,l_z)\leq4$ and $l_w\leq3$ conditions. As
might be seen in the table for binding energy convergence this is
a reasonable choice. When comparing different interaction models
INOY04 results turns to be closest to fully converged (large
basis) result, whereas AV18 has the largest deviation -- but still
of only 150 keV. This is a natural consequence from the fact that
between three selected realistic Hamiltonians INOY04 is the
softest interaction and thus have the fastest convergence both
with respect to PW expansion as well as number of Lagrange-mesh
functions used to describe the radial dependence of FY amplitudes.
On contrary, between three considered interaction models, the AV18
posses the hardest core as well as the strongest tensor
interaction term in $^3SD_2$ wave resulting in relatively slow
convergence.
\begin{table*}[t]
\begin{ruledtabular}
\begin{tabular}{*{4}{c}}
&INOY04 & I-N3LO  & AV18  \\
\hline
 here & -29.09 & -25.24 & -24.08 \\
 large & -29.10 & -25.39     & -24.15\\
 ref.~\cite{Lazauskas_4B,Navratil_fbs07_3bf,Arnas_prc07_4b,Viviani_prc10_alpha}      & -29.11 &  -25.38(1)    &-24.23(1) \\
\hline
\end{tabular}
\end{ruledtabular}
\caption{ \label{tab:he4_be} Binding energies of $^4$He ground
state calculated with basis limitations used in this work (here),
taking the same PW limitation but considerably much larger size of
Lagrange-mesh basis (large), allowing for radial basis
convergence. These results are compared with the literature values
of fully converged calculations.  }
\end{table*}

Though FY equations are formulated for short ranged potentials, in
this work  the repulsive Coulomb interaction, present between the
protons within $^4$He core, is still included. Indeed, as Coulomb
interaction does not intervene in the asymptotic region of the
open scattering channel such a procedure does not impair validity
of FY approach.
\begin{table*}[t]
\begin{ruledtabular}
\begin{tabular}{*{5}{c}}
 &    \multicolumn{4}{c}{$\delta (deg.)$ }\\
 E$_{cm}$  (MeV)               &  \multicolumn{2}{c}{ $J^\pi=\frac{1}{2} ^+$ }  &    \multicolumn{2}{c}{$J^\pi=\frac{3}{2} ^-$}\\
     &  Kohn  & FY     &   Kohn   & FY \\ \hline
0.5  & -22.0  &  -21.3 &    9.10  & 9.34  \\
1.0  & -30.8  &  -30.0 &    38.1  & 38.9   \\
1.5  & -37.5  &  -36.6 &    77.0  & 77.4   \\
2.0  & -43.5  &  -43.1 &    96.9  & 96.5   \\
3.0  & -49.2  &  -48.7 &    107.1 & 105.5  \\
5.0  & -61.6  &  -62.1 &    109.3 & 111.8  \\
7.5  & -71.2  &  -74.1 &    102.1 & 102.0  \\
\hline
\end{tabular}
\end{ruledtabular}
\caption{ \label{tab:n3lo_ph} Calculation of the n-$^4$He
scattering phaseshifts at different energies for
$J^\pi=\frac{1}{2} ^+$ and $\frac{3}{2} ^-$ states and using
I-N3LO potential. Phaseshifts have been calculated employing Kohn
variational principle (Kohn) employing eq.~(\ref{eq_Kohn_vr}) and
from the asymptote of Faddeev-Yakubovsky components (FY) via
eq.~(\ref{eq_Kohn_vr}).}
\end{table*}

In Table~\ref{tab:n3lo_ph}  calculated phaseshifts extracted by
using two different techniques, namely Kohn variational principle
eq.(\ref{eq_Kohn_vr}) and from the asymptote of Faddeev-Yakubovsky
components eq.(\ref{eq_fy_vr}), are presented. These calculations
have been performed for the Hamiltonian based on I-N3LO
NN-iteraction. One may see quite good agreement between the two
methods, difference does not exceed 2\%. As explained in a
previous section, due to approximations used in evaluating
integrals involved for Kohn variational principle, the values
extracted from the asymptote of FY components turns to be more
reliable.

\begin{figure}
\begin{center}
\includegraphics[scale=0.44]{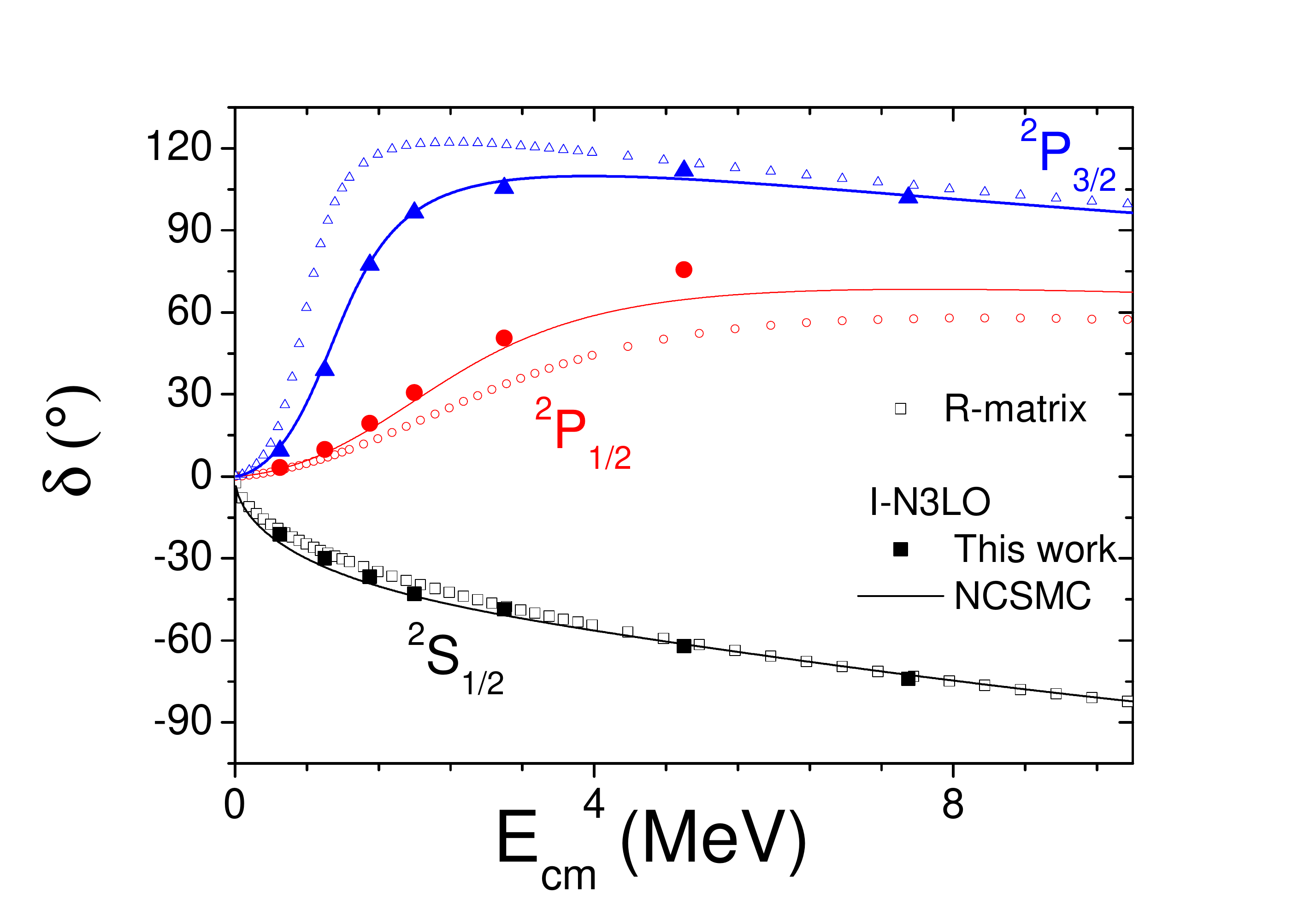}
\normalsize \caption{\label{fig_in3lo_nhe4} Low energy n-$^4$He
scattering phaseshifts calculated for the Hamiltonian based on
I-N3LO nucleon-nucleon interaction. Results of this work (full
symbols) are compared to ones obtained by NCSMC
method~\cite{Navratil_ps16_nhe4} (full lines). Theoretical
calculations are also compared with the phaseshifts  obtained from
the R-matrix analysis
 of the experimental data~\cite{Hale_R_matrix} (open symbols).}
\end{center}
\end{figure}
In Figure~\ref{fig_in3lo_nhe4} $J^\pi=\frac{1}{2}^\pm$ and
$J^\pi=\frac{3}{2} ^-$ angular momenta phaseshifts calculated for
I-N3LO Hamiltonian are compared with the results obtained using
NCSMC technique~\cite{Navratil_ps16_nhe4} as well as with
phaseshifts extracted from experimental data performing R-matrix
analysis~\cite{Hale_R_matrix}. Keeping in mind that both
theoretical calculations -- of this work as well as ones obtained
using NCSMC technique~\cite{Navratil_ps16_nhe4} -- have comparable
numerical accuracy of 1-2$^\circ$  one may signal full agreement
between two
 completely different approaches to solve elastic scattering problem.  In that relates to comparison with the experimental data
 one may also signal nice agreement for the S-wave scattering, dominated by strong Pauli repulsion between projectile
 neutron and ones present within $^4$He target. On contrary description of resonant P-waves is not satisfactory,
 revealing insufficient splitting between  $J^\pi=\frac{1}{2}^-$ and $J^\pi=\frac{3}{2}^-$  waves.

\begin{figure}
\begin{center}
\includegraphics[scale=0.44]{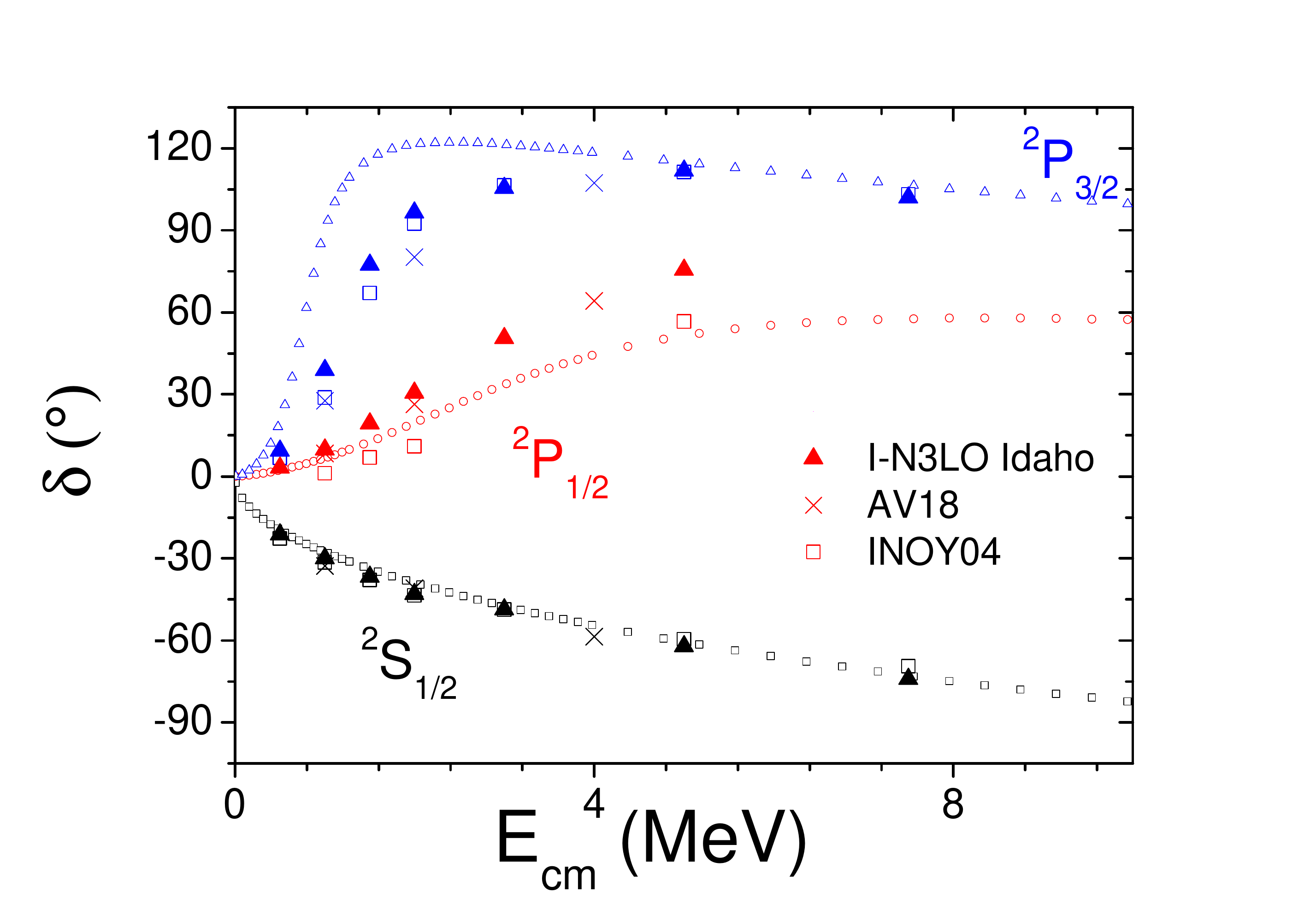}
\normalsize \caption{\label{fig_models_nhe4} Comparison of low
energy n-$^4$He  scattering phaseshifts calculated for different
realistic NN interaction Hamiltonians: I-N3LO (full triangles),
AV18 (crosses) and INOY04 (open squares).  Theoretical
calculations are also compared with the phaseshifts obtained from
the R-matrix analysis
 of the experimental data~\cite{Hale_R_matrix} (small open symbols).}
\end{center}
\end{figure}

 In Figure~\ref{fig_models_nhe4} aforementioned I-N3LO Hamiltonian results are compared with those obtained for INOY04 and AV18 Hamiltonians.
 All the models describe well S-wave phaseshifts, indicating that description of these waves are model independent.
 On contrary different model predictions deviate in describing resonant scattering in P-waves. Both INOY04 and AV18 models
 lack attraction in $J^\pi=\frac{3}{2}^-$ wave, predicting much flatter resonant structure than provided by R-matrix analysis
 of the experimental data~\cite{Hale_R_matrix} or even when compared to I-N3LO results.
  As a consequence splitting between two P-waves  for  AV18 model is smaller than for I-N3LO, which indicates necessity of
  weaker effective spin-orbit interaction for AV18 model than for I-N3LO.
  For INOY04 interaction simple splitting of P-waves
 is not enough to account for the experimental data, for this model P-wave needs to be  much more attractive in overall.
 Worths noting that very similar observations has been made when studying neutron scattering on $^3$H
 nucleus~\cite{Lazauskas_4B,PhysRevLett.111.172302,bench_nH3_pHe3}:
INOY04 model
 lacks strongly attraction in P-waves, AV18 also provides flatter than required resonant structures of $^4$H nucleus,
  whereas I-N3LO model provides the
 best description of the experimental data. This feature indicates on possible correlation between P-wave states
 of $^5$He and $^4$H (or its isospin symmetry partner  $^4$Li  nucleus).

It is well accepted that proper description of the nuclear systems
requires presence of three-nucleon force. Modern models of
three-nucleon forces provide an extra-binding for symmetric
nuclei, like ground state of $^4$He, but also are able to invoke
more attraction for P-wave states. It is demonstrated
in~\cite{PhysRevLett.111.172302} for n-$^3$H scattering and
in~\cite{Navratil_ps16_nhe4} for n-$^4$He case, that
implementation of local 3NF force, developed up to N2LO terms
in~\cite{Navratil_fbs07_3bf}, in conjunction with I-N3LO
interaction improves significantly description of P-wave resonant
states. Very similar effects are observed when implementing
phenomenological IL2 or IL7 three-nucleon
forces~\cite{IL_3NF_forces,Pieper_IL7} in conjunction with AV18
NN-interaction~\cite{PhysRevLett.111.172302,Nollett_prc07_nhe4}.
In this context case of INOY04 interaction turns to be quite
nontrivial. On one hand this model provides proper binding
energies of trinucleon(s), however it slightly overbinds $^4$He
ground state by about 800 keV~\cite{Lazauskas_4B}. More
importantly this model systematically underestimates mean square
radii of the light nuclei~\cite{PhysRevC.79.021303} resulting
large saturation densities of the symmetric nuclear
matter~\cite{Numat_baldo}. Finally, this model is unable to
provide sufficiently attraction for P-wave structures. Therefore
it should be highly nontrivial to correct all these defects by a
simple model of 3NF. INOY04 would require 3NF, which is strongly
repulsive at the origin in order to correct nuclear radii as well
as saturation properties of the nuclear matter. On the other hand
it would need some attraction in periphery with little effect on
symmetric nuclei, whereas providing strong attraction for P-wave
structures.


\section{Conclusion}

In the present paper the first solution of five-body
Faddeev-Yakubovsky equations is presented, when describing neutron
elastic  scattering on $^4$He. The developed numerical method uses
no uncontrolled approximations, is numerically very efficient and
includes very large number of partial waves. These developments
allows description of five-nucleon systems considering realistic
nuclear Hamiltonians.

Three realistic nucleon-nucleon Hamiltonians have been tested,
namely INOY04, I-N3LO and AV18. All of the considered models
provide accurate description of low energy n-$^4$He scattering in
S-wave, dominated by strong Pauli repulsion. On contrary model
predictions deviate from the phaseshifts derived from the
experimental data for resonant P-wave scattering. Very similar
effects have been observed when studying n-$^3$H and p-$^3$H
scattering
in~\cite{Lazauskas_4B,PhysRevLett.111.172302,bench_nH3_pHe3},
which indicates on the possible existence of strong correlations
between four and five nucleon sector.

\textit{Acknowledgment.} This  work  was  granted  access  to  the
HPC  resources  of TGCC/IDRIS under  the allocation
2016-x2016056006  made by GENCI (Grand Equipement National de
Calcul Intensif).

\bibliography{Bib_Rimas}

\end{document}